\documentstyle[epsfig,onecolumn]{MN}
\input{epsf}

   \newcommand{\bmin}[1]{\begin{minipage}[b]{#1\linewidth}}
   \newcommand{\emin}{\end{minipage}}

\begin{document}

\def\etal{{\it et al.~\/}}
\def\cf{{\it cf.\/}}
\def\ie{{\it i.e.~\/}}
\def\eg{{\it e.g.\/}}
\def\be{\begin{equation}}
\def\ee{\end{equation}}
\def\ba{\begin{eqnarray}}
\def\ea{\end{eqnarray}}
\def\no{\noindent}
\def\lya{Ly$\alpha$~\/}
\def\msun{M_\odot}
\def\tcmb{T_{CMB}}
\def\tcmb0{T_{CMB}^0}
\def\ltsima{$\; \buildrel < \over \sim \;$}
\def\simlt{\lower.5ex\hbox{\ltsima}}
\def\gtsima{$\; \buildrel > \over \sim \;$}
\def\simgt{\lower.5ex\hbox{\gtsima}}

\title[SN driven Parker instability]{The Influence of the Gravitational Acceleration on the Supernova-Driven 
Parker Instability.} 
\author[A. Steinacker \& Yu. Shchekinov]
       {Adriane Steinacker$^{1,2}$ and Yuri A. 
	   Shchekinov$^{1,3}$\\
	   $^1$Astronomisches Institut, Ruhr-Universit\"at Bochum, D-44780 Bochum, Germany
	   \\ E--mail: adriane@duras.arc.nasa.gov\\
           $^2$NASA Ames Research Center, Moffet Field, CA 94035, USA\\
	   $^3$Department of Physics, Rostov State University,
	   344090 Rostov on Don, Russia 
	   \\ E--mail:  yus@rsuss1.rnd.runnet.ru} 
\date{accepted}
\pubyear{2001}

\maketitle
\label{firstpage}
\begin{abstract}

Within a framework of 2D magnetohydrodynamic (MHD) simulations, we explore the dynamical regimes initiated by a 
supernova explosion in a magnetized stratified interstellar medium (ISM). 
We concentrate on the
formation of large scale magnetic structures and outflows connected with the 
Parker instability. For the sake of simplicity we show only models with a fixed 
explosion energy corresponding to a single SN occuring in 
host galaxies with different fixed values of the gravitational acceleration $g$ and
different ratios of specific heats.
We show that in general 
depending on these two parameters, three different regimes are possible: 
{\it a)} a slowly growing Parker instability on time scales much longer
 than the galactic 
rotation period for small $g$, {\it b)} the Parker instability growing
at roughly the rotation period, which for ratios of specific heats larger than
1 is accompanied by 
an outflow resulting from the explosion for intermediate $g$, and {\it c)} a rapidly growing
instability and a strong blowout 
flow for large $g$. By means of numerical simulations and analytical estimates we show that 
the explosion energy and gravitational acceleration which separate the three regimes 
scale as $Eg^2\sim $const in the 2D case. We expect that in the 3D case this scaling law is 
$Eg^3\sim $const. Our simulations demonstrate furthermore, that a single SN explosion
can lead to the growth of multiple Parker loops in the disc and large scale
magnetic field loops in the halo, extending over 2-3 kpc horizontally and up to
3 kpc vertically above the midplane of the disc. 

\end{abstract}
\section{Introduction}

Recent observations show the presence of dust in extended haloes of spiral 
galaxies which in some cases reaches out up to distances of around 2 kpc 
above the galactic plane (Howk \& Savage 1997). 
The dusty gas clouds 
can account for 10 \% of the total gas mass in the disc. This finding rises the central question of how 
this material is
lifted up to the observed heights. 
There are arguments against the classical scenario of SN explosions as an efficient mechanism for
the transport of material above the disc.
An essential requirement to these 
mechanisms is that they must be soft enough in order to allow the dust grains to survive 
when being spread over such large scales (a more detailed discussion can be found in 
Howk \& Savage 1997). Finally, the structure of the interstellar medium
in the regions of interest may play a role in the transport mechanisms. This structure, in turn,
is determined by the dynamical processes which are taking place in the galactic plane.
On the one hand, the large scale supershells and holes observed in the Milky Way and in nearby face-on
galaxies (Heiles, 1984, Brinks \& Bajaja 1986, Puche \etal 1992,
Kim S. \etal 1999, Walter \& Brinks 1999),
the extended H$_\alpha$ and dust haloes observed in
nearby edge-on galaxies (Dettmar 1992, Sofue \etal 1994, Howk \& Savage 1999, 
T\"ullmann \& Dettmar 2000), and the large scale
gas outflows in dwarf galaxies (Meurer \etal 1992, Martin 1996), are commonly thought to be
connected with coherent supernova explosions. On the other hand the large-scale magnetic loop structures
observed in edge-on galaxies (Dettmar, private communication), the rising and falling gas motions above
spiral arms (Sofue and Tosa 1974), the loops observed in Orion 
(and towards the galactic centre
(Appenzeller 1974, Sofue 1976, Scalo 1985, Sofue and Handa 1984)
and the radio lobes observed in NGC 3079 and NGC 4388 (Duric \etal 1983, 
Hummel \etal 1983 ) are thought to be produced by the Parker instability. 
The Parker instability is also a possible
mechanism for transporting material out of the disc since 
at nonlinear stages it entraines a large amount of gas on kpc-scales.

The stability of a gaseous configuration partly supported by a parallel stratified  magnetic field
in the presence of
a gravitational field perpendicular to the direction of the magnetic field 
was first studied quantitatively by Newcomb (1961). He derived the 
stability criteria and the corresponding growth rates and 
showed that the instability criteria are generalizations of the
Schwarzschild criterion.  
In a series of papers, Parker (1966, 1967, 1979) has discussed the instability
in the context of the Galactic disc, based on an equilibrium model with a constant gravitational
acceleration, a magnetic field parallel to the galactic plane and stratified in vertical direction,
and equipartition between
magnetic, thermal and cosmic ray pressures. He found the system to be unstable if the adiabatic
index is smaller than a given critical value, which depends on whether the wave vectors of the perturbations
are 2- or 3-dimensional and provided the magnetic field is large enough. 

Over the past two decades, the details of the Parker instability have  been filled in by means of
analytical stability analyses and numerical simulations. 
Additional effects and more physics (such as the inclusion of the rotation state
of the galactic disc, the spatial dependence of the gravitational acceleration, and different types
of perturbations) have been included in both analytical and numerical investigations. The interested
reader is referred to review papers such as those of Mouschovias (1996) 
and Shibata (1996).
Here, we will restrict ourselves to the question of whether, and 
under which conditions
the Parker instability can lead to the formation of the observed looplike structures and gas motions
in galaxies, and how it evolves under strongly nonlinear perturbations such as SN explosions.  
Our studies are restricted to 2-D simulations, and we discuss in section 5 some of the uncertainty
in our results arising from this restriction.

A realistic model describing this kind of environment has first been presented by Kamaya \etal (1996).
They account for both the warm gaseous disc and the hot 
halo. They used a two-temperature equilibrium
model in their 2D numerical simulations of the Parker instability triggered by
a supernova explosion taking place in the galactic plane. Under realistic conditions, 
this
kind of perturbation is most likely
to be the reason for triggering the instability, rather then the sinusoidal perturbations which
are commonly assumed. Based on this model, they studied the effect of two values of
the ratio of specific heats and different energy input for the explosion on the instability,
but did not discuss the influence of different values of the gravitational acceleration
on the growth of the instability (Horiuchi \etal 1988, Matsumoto \etal 1988, 
Giz \& Shu 1993,
Kamaya \etal 1997, King, Hong \& Ryu 1997).
The most recent results on the influence of a spatially dependent
gravitational acceleration on the evolution of the Parker instability has been presented by Kim \& Hong (1998) and Kim \etal
(2000). The initial vertical density and gravitational acceleration distributions are chosen such that
the system is in equilibrium under the assumption of isothermality. Most importantly, the gravitational
acceleration reproduces the profile which can be derived from observations of the spatial density distribution
and velocity-distance data (Oort 1960, Bahcall 1984, Kuijken \& Gilmore 1989), 
with $g=4\times 10^{-9}$ cm s$^{-2}$ at a height
of 500 pc above the disc midplane.  

The gravitational acceleration is a parameter that can vary widely
from galaxy to galaxy. 
It is known (van der Kruit \& 
Searl 1981) 
that all disc galaxies have approximately equal thickness of stellar discs,
while the velocity dispersion of stars
varies in a 
wide range (Bottema 1995), and as soon as they are self-gravitating 
the gravitational acceleration must vary as $\sigma^{-2}$. 
Additional evidences can be found in observations of HI in nearby dwarf 
galaxies, where the observed HI scale heights differ considerably from 
galaxy to galaxy: 625 pc for Holmberg II galaxy (Puche \etal 1992), 
350 pc for IC 2574 (Walter \& Brinks 1999), 180 pc for the 
Large Magellanic Cloud (LMC) (Kim S. \etal 1999)
-- while the velocity dispersion of gas varies in a very 
narrow range around $\sigma_g\simeq 8$ km s$^{-1}$.  
On the other hand, the growth rate of the Parker instability and 
the critical wavelength for instability are inversely proportional to
the gravitational acceleration, the latter being important with regards to the perturbations applied
to trigger the instability. Thus, in order to understand the vertical distribution of gas and magnetic fields in
the haloes of galaxies, the evolution of large scale MHD perturbations in the ISM  must be studied for different
values of the gravitational potential. 
  
A SN explosion presents a nonlinear time and space dependent perturbation. 
Its effect on the growth of the instability
can be different from the usual sinusoidal perturbations and depends
essentially on the dynamics of the explosion. 
The energy input from SNe explosions (normally clustered in OB associations) is recognized
to play a dominant role in structuring the ISM and its overal dynamics in
galaxies on scales up to tens of kpc. 
The standard scenario for producing large scale
motions by SNe involves the formation of a superbubble blown out into the gaseous halo
or widely expanding within the disc (Tomisaka \& Ikeuchi 1986, Tenorio-Tagle \etal 1987,
Mac Low \etal 1989, Palou\v s \etal 1990, Mac Low \& Ferrara 
1999). When a superbubble expands into a non-magnetized
ISM, the condition for a blow-out to occur requires the expansion to remain
supersonic until it reaches a vertical distance of about two to three scale heights of the gas
distribution (Kovalenko \& Shchekinov 1985, Mac Low \& McCray 1988). However, in the presence
of a magnetic field the situation changes crucially. Bernstein \& Kulsrud (1965) and Kulsrud
\etal (1965) first addressed the question of how a spatially uniform magnetic field can
affect the dynamics of a SN explosion. Subsequent studies by 
Giuliani (1982) and Ferri\'ere \etal (1991) have contributed to the understanding of different aspects of
the dynamics of a SN remnant in a uniform magnetic field. The main conclusion that can be drawn from
these studies
is that the magnetic pressure supresses the expansion of a remnant perpendicular to the field 
lines, and thus one can expect that in the presence of magnetic field the energy released
by SN explosions is confined in the galactic disc. Tomisaka (1990) concluded from his
numerical study that a magnetic field with a standard strength $B_0=5~\mu$G provides
confinement of the bubbles with typical mechanical luminosity of $L\sim 3\times 10^{37}$ erg
s$^{-1}$. Quite recently, Tomisaka (1998) has
described numerically the dynamics of a superbubble in a vertically 
stratified magnetic field,
specifically addressing the question of whether a superbubble is blown out or remains confined
in the disc.  He concludes that despite the strong influence of the magnetic tension
perpendicular to the field lines, the expansion of superbubbles in 
vertical direction 
is supressed much less than previously assumed and is kinematically similar 
to the hydrodynamical case.

One of the most important feature in the dynamics of superbubbles in a magnetized ISM is that
as a result of the magnetic field compression by the shock front, magnetosonic waves are driven ahead of
the front (Mineshige et al. 1993).
The separation between the magnetosonic wave and the shock front depends on the ratio of the Alfv\'en
velocity to the shock velocity, and increases when the shock is slowing down. Thus, at late
stages in the expansion of a superbubble the magnetosonic front can disturb the magnetic field and gas
on large scales, much larger than the radius of a bubble, which in turn can initiate the
Parker instability when the characteristic radius of the magnetosonic front is comparable to
the
critical wavelength. Once this condition is fulfilled, the flow associated with the Parker
instability drags out the material driven by the superbubble thus favouring its vertical expansion.

In this paper we will concentrate on the role of the gravitational acceleration 
and the ratio of specific heats on the Parker instability 
triggered by a point explosion with emphasis on the possible interconnection between the supernova
explosion and the Parker instability by using {\it a)} 
an isothermal model and {\it b)} a two-temperature model
accounting for both the galactic disc and the halo. In the framework of 2D MHD,
we will show that in general there are three different regimes depending on 
the gravitational acceleration in the disc and on the ratio of specific
heats: 1) When the gravitational acceleration is 
below a certain value, a SN explosion initiates a slowly growing Parker instability 
on time scales much larger than the galactic rotation period;
2) For large values of the gravitational acceleration 
the explosion drives a shock wave which initiates a rapidly growing Parker 
instability with growth rates comparable to typical values of the galactic free-fall
time; 3) For $g \geq 4.5 \times 10^{-9}$ cm s$^{-2}$, and provided the ratio
of specific heats is larger than the isothermal value, 
the Parker instability can amplify the blowout flow -- the resulting 
magnetic structures and gas dynamics are a product of the dynamical interaction 
of these two processes. 
In all cases we start with the most favourable adiabatic index for the evolution of both processes
(close to isothermal). 

The paper is organized as follows: in Section 2 we describe 
our initial model and numerical method, 
in Section 3 we qualitatively analyse the possible 
regimes of the Parker instability by means of energy estimates, in Section 4
we present our results and we close in 
Section 5 with the discussion of the results and their possible astrophysical implications. 
A summary of the results is presented in Section 6.  

\section{Initial model and numerical method }
The dynamics of the investigated system is described whithin the framework of ideal MHD
simulations, by assuming
the gravitational acceleration to be constant. This is a simplifying assumption, which we only use
in order to keep track of the behaviour of the system, and in favour of other important assumptions
(\eg to account for the interaction between disc and halo), work with a spatially dependent
gravitational acceleration being already underway. We also assume that the thermal and the magnetic pressure
are initially equal, \ie the plasma beta $\beta=P_{th}/P_{mag}$ is 1, where $P_{th}$ is the thermal gas pressure
and $P_{mag}=B^2/(8\pi)$ is the magnetic pressure, and neglect the effect of the cosmic rays.
The simulations have been performed using the Nirvana finite difference code (Ziegler 1995,
Ziegler et al. 1996), which is based on the same numerical techniques as ZEUS (Stone and Norman
1992). 
With the above assumptions, the governing equations are:
\begin{equation}
\frac{D \rho}{D t}+\rho \nabla \cdot {\bf v}=0, 
\end{equation}
\begin{equation}
\rho\frac{D {\bf v}}{D t}=-\nabla p -\rho {\bf g} + 
\frac{1}{4\pi}(\nabla \times {\bf B}) \times {\bf B},
\end{equation}
\begin{equation}
\rho \frac{D }{Dt} \left( \frac{e}{\rho} \right)=-p\nabla {\bf v}, 
\end{equation}
here
$ {D }/{D t} \equiv {\partial }/{\partial t} + {\bf v}\cdot \nabla$
denotes the comoving derivative.
\begin{equation}
\frac{\partial {\bf B}}{\partial t}=\nabla \times ({\bf v} \times {\bf B}).
\end{equation}
The system is closed by a polytropic equation of state 
\begin{equation}
 p = \kappa \rho^\gamma,
\end{equation}
where all the quantities have their usual meanings.
The equations are computed in two dimensions in cartesian coordinates, with 
$x$ being the longitudinal direction parallel to magnetic field lines. In this paper 
we restrict ourselves to a constant gravitational acceleration $g$. 
Since the perturbation is already nonlinear, space- and time-dependent, and is therefore
more complicated than the usually assumed non-localized linear perturbation, 
we can obtain a
better understanding of the qualitative dynamics of the system with regards to basic assumptions, 
without contaminating effects from spatially varying $g$. Moreover, it is worth 
stressing that the gravitational acceleration varies on a scale-height much smaller than the 
critical wavelength for the Parker instability -- for example, in the Milky Way 
galaxy, $g$ increases from $g=0$ to $g=3\times 10^{-9}$ cm s$^{-2}$ when $z$ varies 
from $z=0$ to $z\simeq 200$ pc, and at $z\simeq 600$ pc it reaches $g=6\times 
10^{-9}$ cm s$^{-2}$ (Kuijken \& Gilmore 1989). Therefore,
the perturbations of relevant wavelengths, should be largely unaffected by
this variation. 

Our simulations are starting from either one of the following models, which are based on the assumption
for vertical hydrostatic equilibrium:\\
{\it i)} An isothermal model which accounts for the disc only. In this case, 
the density 
distribution is an exponentially decreasing function of $z$:
\begin{equation}
\rho(z)=\rho_0 \exp(-z/H),
\end{equation}
with the scale height $H=(1+\beta^{-1})c_s^2/g$, where $c_s$ is the isothermal sound speed.
For all the runs we have assumed a temperature $T_d=10^4K$ and a midplane density of the disc of
$n_0=1$ cm$^{-3}$. Three different values for the gravitational acceleration, 
$g=3, 4.5$ and $6\times 10^{-9}$ cm s$^{-2}$ have been chosen.
The parameters for the different runs are summarised
in table 1, which also contains the corresponding values for the scale-height and the critical
wavelength $\lambda_{c}=4\pi H \left[\beta\gamma/(2(1+\beta-\beta\gamma)(1+\beta)
-\beta\gamma)\right]^{1/2}$
(Mouschovias 1996). The wavelength of maximum growth for $\beta\sim\gamma
\sim 1$ is $\lambda_m=1.8 \lambda_c$ for the instability.\\
\begin{table*}
\caption{Characteristics of the one-temperature models}
\begin{center}
\begin{tabular}{rrrrrr}
\hline
{\rm run}  & {\rm g}           & {\rm $\gamma$} & {\rm H}     & {\rm $\lambda_c$} & {\rm $\lambda_{m}$} \\
$\,$       & {\rm (cm/s$^2$)}  &  $\,$          & {\rm (pc)}  &  {\rm (pc)}       & {\rm (pc)} \\ 
\hline 
1Tg3       & 3$\cdot10^{-9}$   & 1.05           &180          & 1307              &  2353 \\
1Tg4.5a    & 4.5$\cdot10^{-9}$ & 1.05           &126          &  871              &  1568 \\
1Tg4.5b    & 4.5$\cdot10^{-9}$ & 1.4           &126          & 1867              &  3361 \\
1Tg6       & 6$\cdot10^{-9}$   & 1.05           &94.5         &  653              &  1175 \\
\hline 
\end{tabular}
\end{center}
\end{table*}
\noindent
The computational domain ranges from $z_{\rm min}=0$ to $z_{\rm max}=2100$ 
pc and $x_{\rm min}=0$
to $x_{\rm max}=3600$ pc. The total number of grid-zones is 
$N_x \times N_z=242 \times 142$,
with a grid spacing of $\Delta z=15$ pc.\\
{\it ii)} A two-temperature model, accounting for both, the disc and its halo. The underlying model
corresponds to the one proposed by Kamaya \etal (1996). The density-distribution is calculated
from a prescribed temperature profile: 
\begin{equation}
T(z)=T_d+\frac{(T_c-T_d)}{2} \left[ \tanh\left(\frac{\mid z \mid-z_c}{w_{tr}}\right)+1\right],
\end{equation}
where $w_{tr}$ 
is the thickness of a transition layer between disc and corona, 
$T_c$ is the coronal temperature, and 
for $z_c\gg w_{tr}$ (in our case $z_c=1$ kpc and $w_{tr}=4\Delta z$) 
$T_d$ can be identified with the midplane temperature of the disc. 
The above profile leads to the following density
distribution:
\be
\rho(z)=\rho_0 \frac{T_d}{T(z)} \exp\left[-{\mu I(z)\over 2 \Re g }\right],
\ee
where $\mu=1$ is the molecular mass of the gas in the disc, $\Re$ is the gas constant, and $I(z)$ is 
a function of $T_c, T_d, w_{tr}$ and the height of the disc $z_c$ which
results from the integration. The coronal temperature is  
$T_c=2.5 \times 10^5$K, and the temperature in 
the disc is $T_d=10^4$K. The rather low value of the temperature in
the halo is chosen in order to avoid
strong density and temperature jumps, which lead to numerical errors on
the interface between disc and halo due to averaging of the physical
quantities. 
However, this is a standard 
assumption in simulations of the Parker instability in two temperature models 
(Shibata \etal 1989, Matsumoto \etal 1993, Kamaya \etal 1996). 
In the real ISM, where radiative cooling can be important on relevant 
time scales, this temperature corresponds to the peak of the cooling function, 
and therefore a larger temperature has to be chosen in order to weaken the
effects of radiative 
losses. Shibata \etal
(1989) have shown that the nonlinear dynamics of the instability initiated by a
non-localized sinusoidal perturbation depend only weakly on the ratio
$T_c/T_d$. One can thus expect that for the SN driven Parker instability, the 
results obtained for $T_c=2.5 \times 10^5$K provide us with a qualitatively correct 
understanding for even larger coronal temperatures. 
The runs performed with this model are summarized in Table 2. We used three different values
for the gravitational acceleration, and four values for the ratio of specific heats $\gamma$.\\
\begin{table*}
\caption{Characteristics of the two-temperature models}
\begin{center}
\begin{tabular}{rrrrrr}
\hline
{\rm run}   & {\rm g}           & {\rm $\gamma$} & {\rm H}     & {\rm $\lambda_c$} & {\rm $\lambda_{m}$} \\
$\,$        & {\rm (cm/s$^2$)}  &  $\,$          & {\rm (pc)}  &  {\rm (pc)}       & {\rm (pc)} \\
\hline
2Tg3k1.05   & 3$\cdot10^{-9}$   & 1.05           &180          & 1307              &  2353 \\
2Tg3k1.4    &                   & 1.4            &             & 2675              &  4815 \\
2Tg3k1.75   &                   & 1.75           &             &  ---              &   --- \\
\hline
2Tg4.5k1.05 & 4.5$\cdot10^{-9}$ & 1.05           &126          &  871              &  1568 \\
2Tg4.5k1.4  &                   & 1.4            &             & 1867              &  3361 \\
2Tg4.5k1.6  &                   & 1.6            &             &  ---              &   --- \\
2Tg4.5k1.75 &                   & 1.75           &             &  ---              &   --- \\
\hline
2Tg6k1.05   & 6$\cdot10^{-9}$   & 1.05           & 94.5        &  653              &  1175 \\
2Tg6k1.4    &                   & 1.4            &             &  1404             &  2528 \\
2Tg6k1.6    &                   & 1.6            &             &  ---              &  ---  \\
\hline
\end{tabular}
\end{center}
\end{table*}
The computational domain ranges from $z_{\rm min}=0$ to 
$z_{\rm max}=5400$ pc and from $x_{\rm min}=0$
to $x_{\rm max}=3600$ pc. The total number of grid-zones is 
$N_x \times N_z=242 \times 362$, which results in
a grid spacing of $\Delta x=\Delta z=15$ pc. The point explosion is 
initialized in 
the origin such that a quantity of thermal energy $E=10^{51}$erg and a mass of $M_{\rm expl}=10M_{\odot}$
is concentrated in a cylindrical volume with $r_{\rm cyl}=43$ pc 
and height $Y=1$ kpc. In our simulations the explosion is initialized 
by an instanteneous energy release -- we did not perform computations with a continuous 
energy input with constant energy ejection rate. Such an approach is justified by 
the fact that the Parker instability normally develops on time scales longer than the 
typical time for vigorous evolutionary stages of OB associations.  

\section{Energy requirement}
In order to understand the dynamics of the Parker instability driven by a SN 
explosion, we present qualitative estimates for flows produced in 
a stratified ISM by point explosions. 
For a point explosion in an exponential non-magnetized atmosphere, the energy required
to produce a blow-out can be estimated from the condition that the hot
bubble expands supersonically when its radius $R_s$ reaches $\sim 3H$,
$H$ being the scale height of the gas distribution (Kovalenko
\& Shchekinov, 1985, Mac Low \& McCray, 1988). This can be understood from a simple estimate
of the shock velocity at the upper point of the bubble $z$, which in an atmosphere 
with an exponential density distribution $\rho(z)=\rho_{0}\exp(-z/H)$ and in 3D is:

\be
\label{shvel}
v(z)\sim \left[{E\over \rho_{0}}\right]^{1/2}e^{z/2H}z^{-3/2}.
\ee
This function has a minimum at $z=3H$ (see Ferrara \& Tolstoy 2000). 
Additionally assuming the bubble to be spherical when
$R_s=3H$ and requiring its expansion velocity at this stage be sonic $v_s=\sqrt{\gamma}c_s$ we obtain 

\begin{equation}
\label{shen}
E_{\rm min}^B=\left({5\over 2}\right)^2 27e^{-3}\gamma\rho_{0}c_s^2H^3.
\ee
The minimum energy for a blow out to occur in a magnetized atmosphere can be
estimated by requiring the velocity at $R_s=3H$ to roughly equal the velocity
of the fast magnetosonic 
wave $v_s\sim \sqrt{v_A^2+\gamma c_s^2}$, where $v_A$ is the Alfv\'en velocity.
This leads to 

\begin{eqnarray}
\label{bshen}
E_{\rm min}^B=\left({5\over 2}\right)^2 54e^{-3}(2\beta^{-1}+\gamma)\rho_{0}c_s^2H^3\\
\nonumber
\sim {25\over 2} (2\beta^{-1}+\gamma)\rho_{0}c_s^2H^3.
\ea
Since H is proportional to $c_s^2$ and to $g^{-1}$, eq. (11) shows that
$E_{\rm min}^B$ strongly depends on the sound speed in
the ISM ($\propto c_s^8$), and on the gravitational acceleration in a host galaxy
($\propto g^{-3}$). For
$\rho_{0}=1.67\cdot 10^{-24}$ g cm$^{-3}$, $T=10^4$ K, $\gamma=1.05$ and 
$g=3\times 10^{-9}$ cm s$^{-2}$, 
the resulting minimum energy is $E_{\rm min}^B=4\times 10^{51}$ erg. 
Note, that the value of $E_{\rm min}^B$ required to produce a blow-out in the hot halo 
scales as $E_{\rm min}^B\propto \rho_{\rm h}T_{\rm h}^4$, where $\rho_{\rm h}$ 
and $T_h$ are
the density and temperature of the halo at the disk midplane.
At a density corresponding
to $n_{\rm h}\sim 
0.01$ cm$^{-3}$ and $T_{\rm h}=25\times 10^4$ K the minimum energy is $E_B=3\times 10^{54}$ erg. 
This estimate was obtained by
neglecting radiative losses, \ie assuming adiabatic expansion, and can thus be
considered as a lower limit. This assumption is reasonable. As shock waves
expand into a magnetized medium, radiative losses are not particularly
severe because the post-shock compression is always
weaker in the presence of a magnetic field.

In general, the presence of a magnetic field provides counter-pressure which
tends to weaken the propogation of a shock.
Within a uniform magnetic field a hot bubble elongates 
along the field lines, and for a standard energy input or mechanical luminosity,
remains
confined in the galactic disc (Tomisaka 1990). In a stratified magnetic field with
$\beta=1$, the
situation changes as compared to the non-magnetized case.
The most interesting new dynamical feature here is connected with the
acceleration of the shock front in the vertical direction and
deceleration of the contact discontinuity between the hot bubble
and the shocked ISM gas, so that the vertical size of the hot bubble
itself is smaller than in the non-magnetized medium, while the 
thickness of the layer of shocked gas and the vertical size of
the disturbed zone is larger (Mineshige, Shibata \& Shapiro 1993, Tomisaka
1998).
In these calculations the time and spatial
scales were restricted to $t\sim 10-40$ Myr and $R\sim 0.6-1$ kpc, respectively,
and the main conclusion of whether a SN produced bubble is confined to the disc or
blown out is drawn from the violent early evolutionary stages of the bubble and its
associated shock wave. However, at later stages, even a weak shock (or acoustic)
wave can stimulate the growth of the Parker instability. 

The minimal SN energy
required to produce the Parker flow can be derived from the 
following requirement. The size of the region
occupied by the magnetosonic (or Alfv\'en) wave $R_B$ must grow
larger than the critical
wavelength of the instability $\lambda_c$, and the amplitude of the wave 
must be large enough to allow the instability to grow on time scales
less than the
rotation period $t_R$ of the galaxy. 
In the linear regime, 
the velocity amplitude grows as $v\sim v_0\exp(t/t_P)$, where $v_0$ is the initial 
amplitude, and $t_P\sim H/c_s$ is the Parker growth time. If the velocity
amplitude is to grow to the fast-mode speed by the time $t_i$, then 
$v_0\sim \sqrt{v_A^2+\gamma c_s^2}\exp(-t_i/t_P)$. Typically, a remnant will 
expand for about $50$ Myr, the value that we will for $t_i$, and
$t_P\sim 10$ Myr. Hence, 
we require $v_0\sim 10^{-2}\sqrt{v_A^2+\gamma c_s^2}$. 

In order to estimate the offset time for the Parker instability, we
assume that the shock front moves vertically relative to the contact 
discontinuity with constant velocity equal to the magnetosonic
speed $v_m$ and the discontinuity itself propagates with a velocity that
coincides by the order of magnitude with the shock velocity of
a spherical SN remnant of radius $R_s$ in the Sedov-Taylor phase
$v_d\sim v_s=2R_s/5t$. Thus, vertically, the magnetosonic front reaches
the distance:  
\begin{equation}
R_z\simeq R_s+v_mt=R_s\left[1+{2\over 5}{(v_A^2+\gamma c_s^2)^{1/2}\over v_s}\right],
\ee
where $R_s$ is the radius of the bubble (assumed spherical), and
$v_s$ is the shock velocity. Here the expansion time,
$t=2R_s/5v_s$, for an adiabatic bubble has been substituted. Horizontally, the
shock front propagates with the Alfv\'en velocity and reaches the distance:

\begin{equation}
R_x\simeq R_s+v_A t\simeq R_s\left[1+{2\over 5}{v_A\over v_s}\right].
\ee
We assume that when $2R_B=2\sqrt{R_z^2+R_x^2}=\lambda_c$, the Parker 
instability
begins to grow. The minimum energy can then be obtained by assuming the shock wave
is weak at this stage, \ie $v_s\sim \sqrt{v_A^2+\gamma c_s^2}$. 
For $\beta=1$, this leads to $R_B\simeq 2R_s$, and thus the condition for the Parker instability
to grow is $R_s\simeq \lambda_c/4$, or $R_s\simeq 1.8 H$. 
Substituting this
value into Eq. (9) for $v(z=1.8H)\sim \sqrt{v_A^2+\gamma c_s^2}\exp(-t_R/t_P)$,
we 
obtain
\begin{equation}
\label{magen}
E_{\rm min}^P\sim \left({5\over 2}\right)^2(2\beta^{-1}+\gamma)
\exp(-2t_R/t_P)\rho_{0}c_s^2H^3.
\ee
By comparing (11) and (14) we see
that in the presence of a stratified
magnetic field with $\beta\sim 1$, the explosion energy required to 
initiate the Parker
instability is much less than the explosion energy required to produce a blow out. 
For $t_R\sim 50$ Myr and $t_P\sim 10$ Myr, $E_{\rm min}^P$ can 
be only $\sim 10^{-4}~E_{\rm min}^B$ in the one-temperature model, 
and $\sim 0.1 E_{\rm min}^B$ in a two-temperature model with 
$t_P\sim 50$ Myr in the halo.
This energy can be considered as an upper limit if we realize that even
subsonic (slow magnetosonic) waves can be amplified in an atmosphere with 
a decreasing density profile.
In an exponential atmosphere the velocity amplitude of acoustic waves
increases as (see Lamb 1932),

\begin{equation}
v\propto e^{z/2H}.
\ee
When the Parker instability initiated by magnetosonic perturbations starts to grow, the
amplification can be stronger due to the fact that the flow associated with the instability
removes gas from the upper parts of the forming loop and the perturbation propagates in
a medium with a temporally decreasing density. If we assume a density 
decrease described by $\rho\propto \exp(-t/t_P)$, 
then the velocity amplitude of the magnetosonic wave can grow as

\begin{equation}
v\propto \exp {z\over 2H}+{t\over 2t_P}.
\ee
As a result, a secondary shock wave can build up in response to the subsequent steepening of
the front of the magnetosonic wave. Such secondary shocks appear in the numerical simulations
shown in Section 5.

In the 2D case, the expansion law for a shell produced by a point explosion
is given by

\be
v(z)\simeq {1\over 2}\left[{E\over \rho_{0}}\right]^{1/2}e^{z/2H}z^{-1},
\ee
which leads to the estimate for the minimal blow out energy (equivalent to 
Eq. 9) in the 3D case of:

\be
\label{bshen2}
\varepsilon^B={E^B\over Y}\sim 2(2\beta^{-1}+\gamma)\rho_{0}c_s^2H^2, 
\ee
and to the minimal energy for the Parker instability (equivalent to Eq. 
14 of 

\be
\varepsilon^P={E^P\over Y}\sim 2(2\beta^{-1}+\gamma)\exp(-t_R/t_P)\rho_{0}c_s^2H^2,
\ee
where $\varepsilon$ is the explosion energy per unit length and 
$Y$ is the size of the computational domain in $y$. 

\section{Results}
\subsection{One temperature model}
In order to describe the actual situation in a galactic environment in a realistic
way, one needs to account for the observed difference in temperature between the
disc and the halo. Nevertheless, in order to help fix ideas, we
have performed several runs 
within the framework of an isothermal distribution. We were motivated by the 
fact that most of the previous numerical studies of the Parker instability 
initiated by non-localized (smoothly distributed) 
large scale perturbations have been conducted within the confines of
an isothermal ISM (Basu, Mouschovias \& Paleologou 1997, Kim J. \etal 1999, 
Kim \etal 2000).  
A localized point-like energy release, which is growing in time, should trigger 
the various phases of an unfolding dynamical instability
on different timescales than a non-localized source of energy release.
Therefore, the primary reason for our examination
of isothermal models is to compare the dynamical features of the Parker instability
driven by perturbing sources with a different character of the spatial and temporal
distribution. Unlike the case of a 
non-localized linear perturbation, the SN explosion itself is a nonlinear 
dynamical process, and hence, under appropriate conditions, the two processes 
-- the SN explosion and the Parker instability -- can interact and amplify 
each other depending on the given time scales. Such an interaction can be 
better understood in simple isothermal models
because the sound speed is spatially homogeneous and because the
contaminating features which plague non-homogeneous (two temperature) media 
(see Section 4.2) do not appear. 

The energy input which we apply in order to start the explosion and trigger
the instability (see section 2) does not always 
correspond to a strongly nonlinear perturbation. Its 
relative magnitude can be estimated as the ratio $E/E_{\rm th}$, where $E_{\rm th}=
(\gamma-1)^{-1} k_Bn_0TV$, $V$ being the volume of the perturbed region. For 
example, for $\gamma= 
1.05$ and $V=43^2\times 1000$ pc$^3$, this is only $E/E_{th}\simeq 0.43$.  
Therefore, for $\gamma \sim 1$, the energy input $E=2\cdot10^{51}$ erg does not 
correspond to a SN explosion but rather to a point-like nearly linear 
perturbation -- most of the astrophysical details of a realistic energy input 
from a SN are left aside. It allows us, however, to understand how a Parker 
instability initiated by a weak localized perturbation develops under these 
basic assumptions.

We thus expect that when $\gamma$ is small and the gravitational acceleration 
is less than $\approx 5\cdot10^{-9}$ cm s$^{-2}$), the 
growth rates for the Parker instability are much 
smaller than the existing growth rates for similar models, based on 
non-localized (sinusoidal) perturbations. The reason for this difference is
that a point-like perturbation can initiate the instability only when it has propagated 
to overcritical spatial scales, and since for small $\gamma$ the produced 
perturbation is weak and propagates with sound speed, it takes $t\sim \lambda_c/
c_s$ Myr until the instability can start to grow. Moreover, the magnitude of the 
perturbation when it reaches overcritical scales is very small. 
On the other hand, a non-localized 
overcritical perturbation starts to grow directly at $t=0$.
When $g$ becomes larger $\lambda_c$ becomes smaller and the stratification becomes
stronger. A localized perturbation can therefore propagate more rapidly up to the critical scale.
Additionally, for larger $\gamma$, the perturbation increases and a shock
wave can be produced.
As a result, the instability is growing much
faster, on scales comparable to those obtained for a non-localized perturbation.

\subsubsection{Small $\gamma$, small $g$} 
The simulations performed in this category correspond to model 1Tg3
in table 1.
The energy density contained in the perturbed region is 
$\delta\epsilon=1.2\times 10^{-11}$ erg cm$^{-3}$, while the background thermal 
energy density is 
$\epsilon=2.76\times 10^{-11}$ erg cm$^{-3}$. Therefore, the perturbation is 
only weakly nonlinear with a relative magnitude 
$\delta \epsilon/\epsilon\simeq 0.4$, by a factor of $\simeq 14$ 
less than for a monoatomic gas with $\gamma=5/3$. 
In the initial stages, a sonic shock wave 
forms, which later becomes subsonic  
and propagates as a linear perturbation -- a
magnetosonic wave. 
The estimated time for the perturbation to reach
a height equal to the critical wavelength in this case is about 126 Myr. In
models with a sinusoidal perturbation, the Parker instability at this stage
was already well into the nonlinear regime (see Mouschovias 1996, Matsumoto 
\etal 1988
for comparison). In our case, only after this time has passed, is the instability 
initiated. After 296 Myr, the Parker flow is well developed with maximum
velocities at the footpoints of the loop of about 4$c_s$ and an increase in thermal
pressure at the foot points of the loop of up to one order of magnitude with respect
to the initial values. 
The elevation of magnetic field lines, however, is not very prominent and a saturation is reached
after 296 Myr.

\subsubsection{Small $\gamma$, large $g$}
Model 1Tg4.5a.   
Here the gas scale height is smaller than in the previous case, and therefore the 
shock wave is stronger at given $z$. In addition, 
the critical wavelength is smaller 
in this case, so that the perturbation covers the scale of one critical
wavelength after only 53 Myr.  
The supersonic flow sparks rapid growth of the Parker instability 
because the velocity and magnetic field perturbations are nonlinear: 
$v\simgt c_s$, $\delta B\simgt B_0$. At these stages,
feedback features are observed in the dynamics of the outflow: due to the growing 
Parker instability, gas starts to slide downward along the field lines and the 
density in the upper parts of the expanding shell decreases. This allows for
the hot bubble and the lightened shell to expand rapidly upward. However, 
due to this interaction between the two flows -- the expanding gas and the 
Parker instability --  the amount of mass ejected into the halo is 
smaller in comparison with the models where the instability cannot grow.  
The velocity of both the outflowing gas carried by the magnetic field loops,
and the gas sliding downward along the loops, is strongly supersonic, reaching
velocities of up to $5c_s$ (the plasma beta reaches 0.01 in the middle loop regions
and 100 in the foot point regions, for all z. 

\begin{figure*}
\epsfxsize=10truecm
\epsfysize=10truecm
\centerline{\psfig{figure=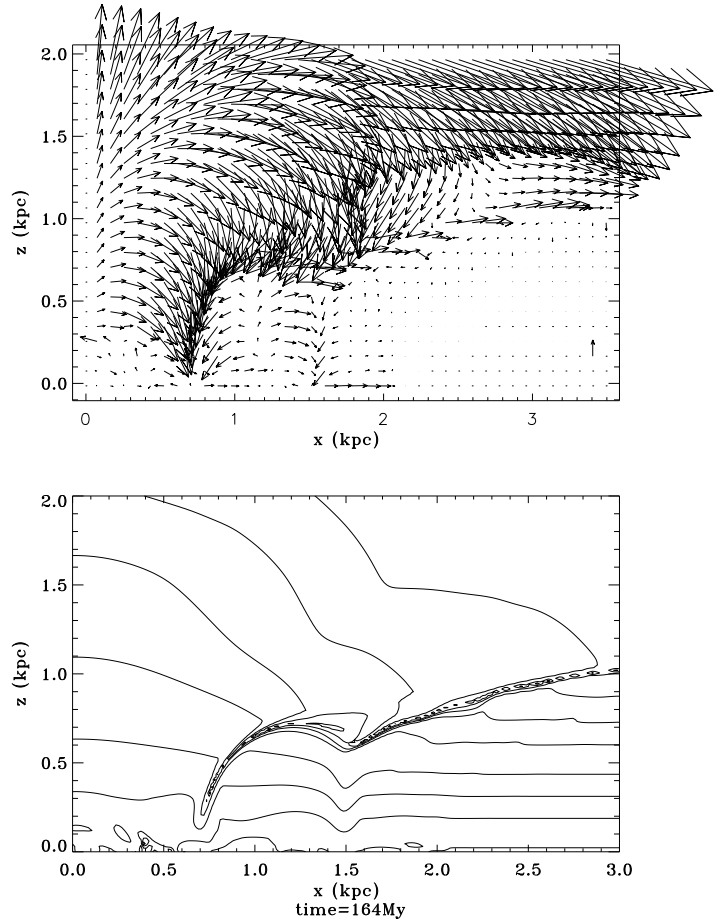,height=120mm,angle=0}}
\caption{\footnotesize{Velocity vectors in the $x-z$-plane (upper panel)
and magnetic field lines (lower panel) for $t=164$ Myr,
$g=6\times 10^{-9}$ cm s$^{-2}$ and $\gamma=1.05$.}}
\end{figure*}

Model 1Tg6. 
Due to the further decrease of the scale height, the amplitude of the 
magnetosonic wave is larger at a given $z$ than in the previous cases, and when 
the wave reaches approximately 2$H$ (at $t\sim 20$ Myr) it becomes supersonic. 
In $t\sim 43$ Myr the perturbation has propagated up to the actual $\lambda_c$ and the Parker 
instability can grow. As seen in Fig. 1 at $t=164$ Myr, the Parker flow is 
already well developed, so that the velocity profiles near the vertical 
symmetry plane show a self-similar behaviour, typical for the advanced stages of the 
Parker instability (Shibata \etal 1989), with the gas being accelerating upward. 
There is also evidence for a small secondary loop centered at $x=1.2$ kpc
with roughly the critical wavelength. 
In the next Section we will see that in the two temperature models, 
the effects of the
self-propagating Parker instability
(the formation of multiple loops) are more pronounced. The velocity vectors are
normalised to the sound speed, which is represented in the lower right corner
of the domain. 
After 164 Myr, the plasma beta has decreased by about two orders of magnitude
with respect to the initial equipartition value. 
At the foot points of the loops, there is evidence
for strong shock waves, built from gas sliding downward with supersonic
velocities. 
These models tend to show that a
larger gravitational acceleration, $g$, results in larger flow velocites for
a given time and height and a stronger decrease of the plasma beta.  

\subsubsection{Large $\gamma$}

Model 1Tg4.5b. 
When $\gamma$ increases, the relative perturbation increases according to 
$\delta \epsilon/ \epsilon \propto (\gamma-1)$ (for $\gamma=1.4$,  
$\delta \epsilon/\epsilon\simeq 3.6$). This means that the perturbation can  
drive a strong shock wave and forms a shell from the very beginning. 
Even though $\lambda_c$ is very large in this case (about three times larger
than for $g=6\times 10^{-9}$ cm s$^{-2}$), the perturbation propagates strongly
supersonically and after $t=40$ Myr, the scale of $\lambda_c$ is reached
and the instability can grow. 
Further expansion leads to acceleration of the 
shock and the postshock flow due to both 
exponential decrease of the density, and evacuation of gas from the upper regions 
of the shell.  
The associated Parker flow is already highly developed. 
Profiles of vertical velocity, in a narrow region close to the vertical symmetry 
plane, show signs of a self-similar behaviour which differs from the linear 
self-similar velocity profiles described by Shibata \etal (1989). We assign this 
difference to the interaction between the expanding shock and the Parker 
instability. By comparing models 1Tg4.5a and 1Tg4.5b, 
we can thus conclude that increasing $\gamma$ while keeping the gravitational acceleration 
constant results in the flow displaying more vigorous dynamics. This is a direct consequence of 
the larger relative amplitude of the perturbation for larger $\gamma$. 
 
\subsection{Two temperature model}

The observed HI density distribution (see Dickey \& Lockman 1990), as well as 
the distribution of X-ray and radio-continuum emission (Kalberla \& Kerp 1998) 
suggests that the ISM is intrinsically multi-component. For our purpose, the ISM 
can be reasonably described by a two temperature distribution with a warm disc 
and a hot overlying corona (see Shibata \etal 1989, Kamaya \etal 1996). 
In such models,
new dynamical effects appear.
On one hand, due to the large sound speed in
the halo, the SN-triggered wave starts to propagate much faster when it
reaches the interface.
The parts of the wave front propagating along the interface therefore disturb
the gas in the disc through associated perturbations of pressure,
and thus, at a given time, the perturbation covers much larger distances
in $x$ than in the isothermal case. In 
general, the development of the Parker instability is favoured, and 
multiple loops are produced in the disc.
On the other hand, due to the much larger scale height of
the hot halo, the instability could saturate when the loops are entering
this medium. 

Kamaya \etal (1996) have obtained the strongest evolution of the SN driven 
Parker instability 
extending deep into the halo in their two temperature model. However, they adopted a 
specific heat ratio of $\gamma=1.05$ for which, as we mentioned above,  
the relative perturbation is rather small and can be considered as weakly 
nonlinear point-like. In order to better understand how the system 
responds to nonlinear perturbations with larger relative magnitude, we have studied
the effect of larger 
$\gamma$ (including $\gamma>\gamma_c$).  
We have perfomed exploratory simulations for a wide range of parameter 
combinations, which are summarized in Table 2.

\begin{figure*}
\epsfxsize=10truecm
\epsfysize=10truecm
\centerline{\psfig{figure=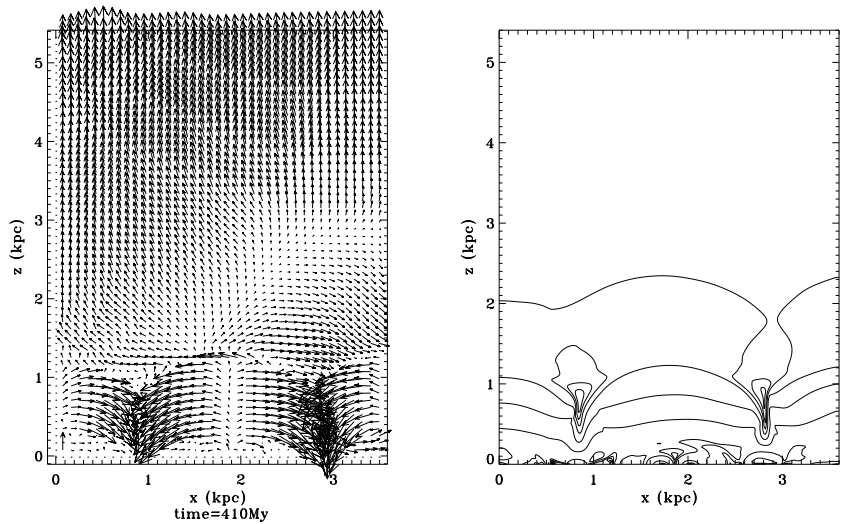,height=120mm,angle=90}}
    \caption{\footnotesize{Velocity vectors in the $x-z$-plane (left panel)
and magnetic field lines (right panel) for $t=410$ Myr,
$g=3 \times 10^{-9}$ cm s$^{-2}$ and $\gamma=1.05$.}}
    \end{figure*}

\begin{figure*}
\epsfxsize=10truecm
\epsfysize=10truecm
\centerline{\psfig{figure=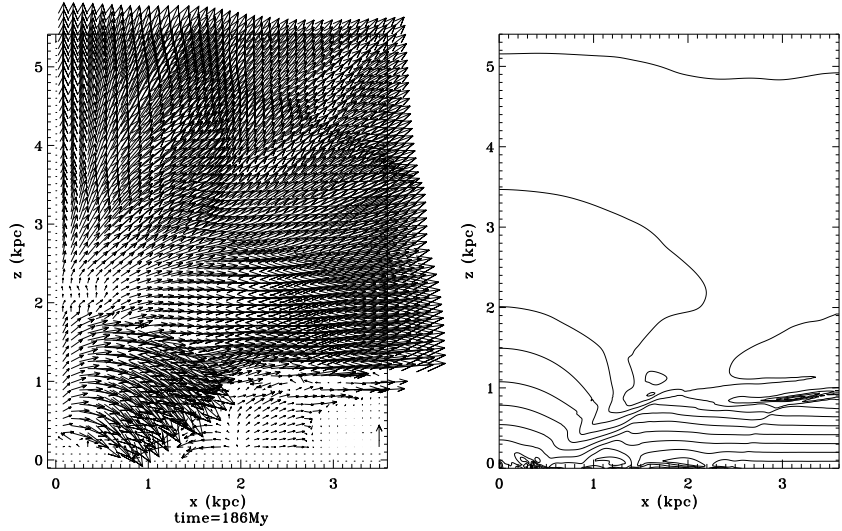,height=120mm,angle=90}}
    \caption{\footnotesize{Velocity vectors in the $x-z$-plane 
and magnetic field lines at
$t=186$ Myr (lower panel) for $g=4.5 \times 10^{-9}$ cm s$^{-2}$
and $\gamma=1.6$.}}
    \end{figure*}

\subsubsection{Low $g$}

Model 2Tg3k1.05. 
In this case, the relative perturbation is weak, and thus 
the explosion front propagates 
with subsonic velocity up to 1-2 scale heights of the disc. Subsequent expansion in an exponential environment is accompanied by the amplification of the wave, which becomes
supersonic when the front reaches a height of about 4H. 
In about 50 Myr, the perturbation has reached the interface between the disc and 
the halo, and in the following $\sim 30$ Myr, it covers a vertical range of 
$\sim 3$ kpc, while horizontally in the disc the critical wavelength is covered
at this time. 
As a consequence, a weakly growing Parker 
instability is initiated with well defined loops building up in about 410 Myr, as
seen in Fig. 2, which shows the velocity vectors and associated magnetic field lines.
\begin{figure*}
\epsfxsize=10truecm
\epsfysize=10truecm
\centerline{\psfig{figure=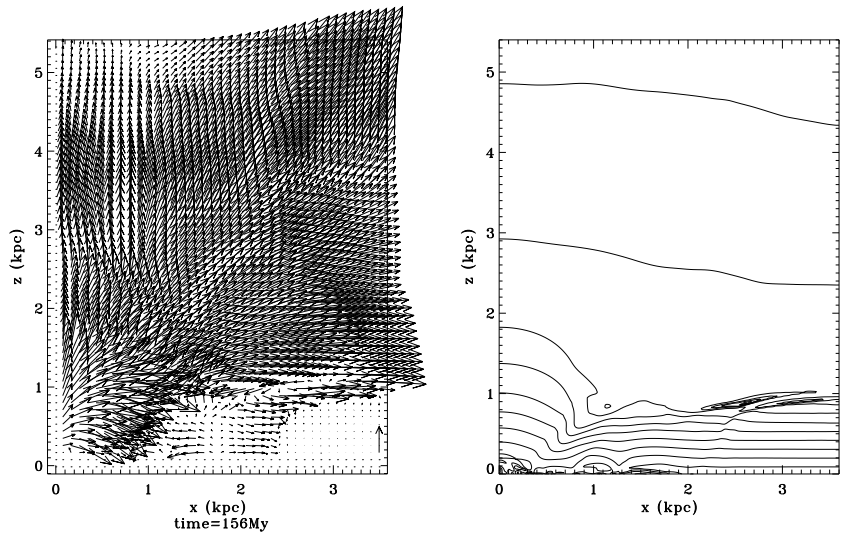,height=120mm,angle=90}}
 \caption{\footnotesize{Velocity vectors (left panel) and magnetic field lines (right panel) at
156 Myr for $g=4.5 \times 10^{-9}$ cm s$^{-2}$
and $\gamma=1.75$.}}
    \end{figure*}
The vertical arrow in the left lower corner of the domain represents
the sound speed, to which the velocity of the flow is related.
While the wavelength of the main loop centered at about 2 kpc can be estimated to be about 2 kpc, which
 is close
to the wavelength of maximum growth (see table 2), the two half-loops that are
also
seen in the picture have smaller wavelength. This difference is a consequence
of the interplay between the applied nonlinear perturbation and the interface.
The perturbation propagating faster in the halo than in the disc is disturbing the interface and is bei
ng transmitted in this indirect way to the disc
driving a propagating Parker instability: when it
covers a horizontal distance comparable to the wavelength of
maximum growth, the
instability forms a loop in shortest time. To the contrary, the half-loop centered
at the origin is initiated directly by the explosion,
and is growing
slower. This is a consequence of the implied nonlinear perturbation.
The density increase in the
valleys relative to the ambient medium for a given $z$ reaches about a factor of 10. 
We want to emphasize, however, that this
growth time is much larger than the typical growth times obtained as a result
of sinusoidal perturbations (see Matsumoto \etal 1988, Mouschovias 1996, Kim 
\etal 2000),
and depending
on the rotation time of the galactic disc at the given distance from its
centre, the instability might become irrelevant on interesting time scales.  
In addition, the elevation of magnetic field loops from the disc into the halo
is not very significant, reaching out to at most 2 kpc.

Model 2Tg3k1.4. As $\gamma$ increases, 
the perturbation also increases as described in the one-temperature models, and the flow 
is more strongly supersonic in both $z$ and $x$. Even though the critical wavelength
is larger in this case,  
it can be reached by the perturbation faster than in the
case of lower $\gamma$. The simulations therefore reflect a 
smaller growth time than in the previous case, while the overall evolution
of the system is very similar, including the formation
of the secondary loops dicussed above.

For $\gamma=1.75$, the Parker 
instability is not growing on relevant time scales, contrary to the case of larger gravitational 
acceleration, where, as discussed in the next section 
loops are still formed.

\begin{figure*}
\epsfxsize=10truecm
\epsfysize=10truecm
\centerline{\psfig{figure=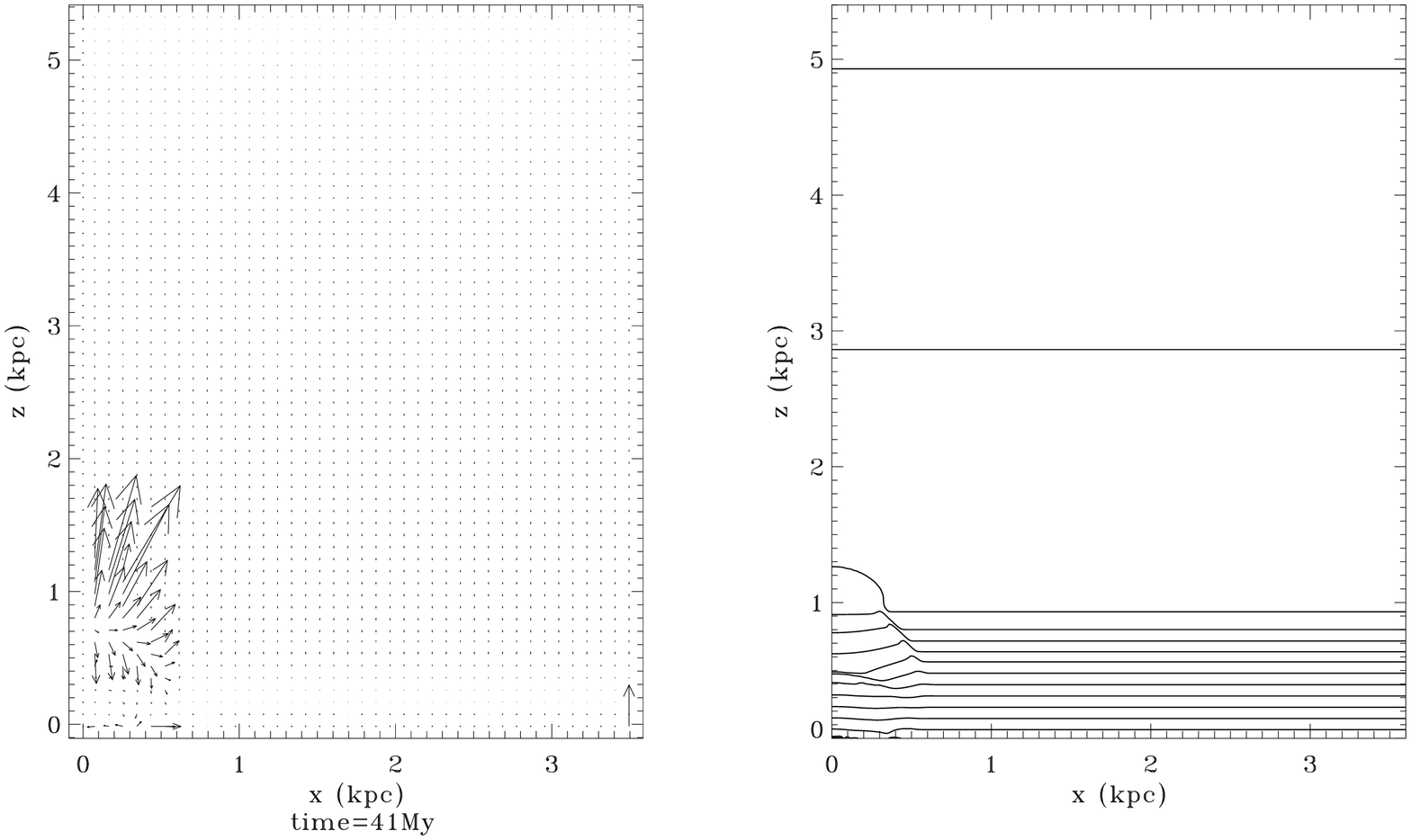,height=120mm,angle=90}}
\medskip
\centerline{\psfig{figure=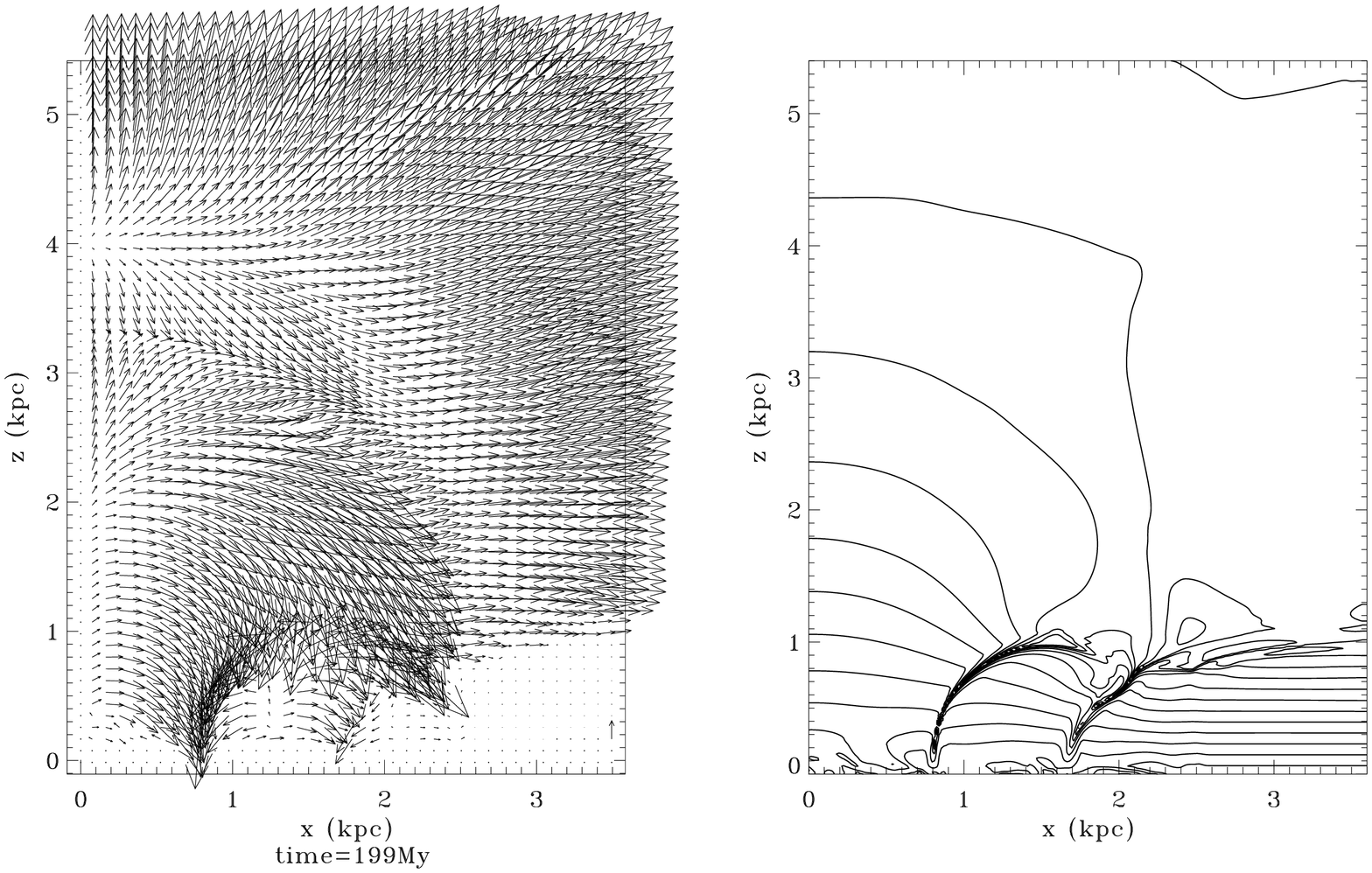,height=120mm,angle=90}}
 \caption{\footnotesize{Velocity vectors and magnetic field lines at
41 Myr (bottom) and 199 Myr (top) for $g=6 \times 10^{-9}$ cm s$^{-2}$
and $\gamma=1.05$.}}
    \end{figure*}

\subsubsection{Medium $g$}

The models with $g=4.5\times 10^{-9}$ cm s$^{-2}$ are characterised by lower scale heights and critical
wavelengths, and 
show faster evolution than for smaller gravitational acceleration as expected from
linear analysis. In model 2Tg4.5k1.05, the 
initial relative perturbation is small, and the dynamics of the stimulated flow is 
similar to that in model 2Tg3k1.05, although more vigorous. The Parker instability 
therefore grows on shorter time scales. However, even for this stronger
gravitational acceleration, significant loops with density enhancement in
the pockets comparable to those obtained in the previous case are built only after
330 Myr. In the simulations with larger 
$\gamma$, models 2Tg4.5k1.4, 2Tg4.5k1.6 and 2Tg4.5k1.75, the initial relative 
amplitude of the perturbation becomes much larger 
so that shock waves form.   
Fig. 3, where we represent the velocity vectors and
magnetic field lines at $t=186$ Myr
for $\gamma=1.6$,
shows the well developed Parker
loops and associated gas flow. A second loop with a smaller wavelength is
visible in this plot.

\begin{figure*}
\epsfxsize=10truecm
\epsfysize=10truecm
\centerline{\psfig{figure=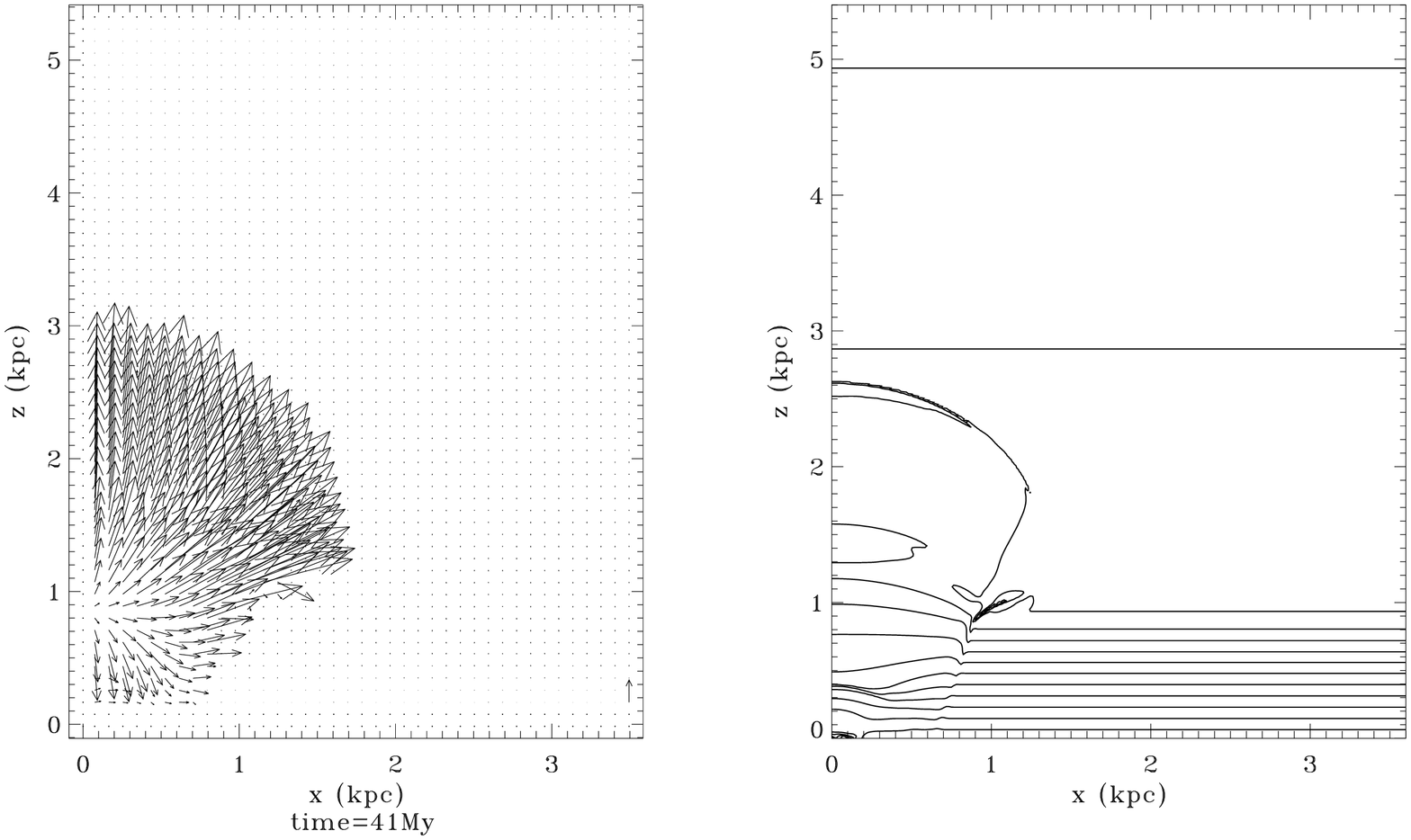,height=120mm,angle=90}}
\medskip
\centerline{\psfig{figure=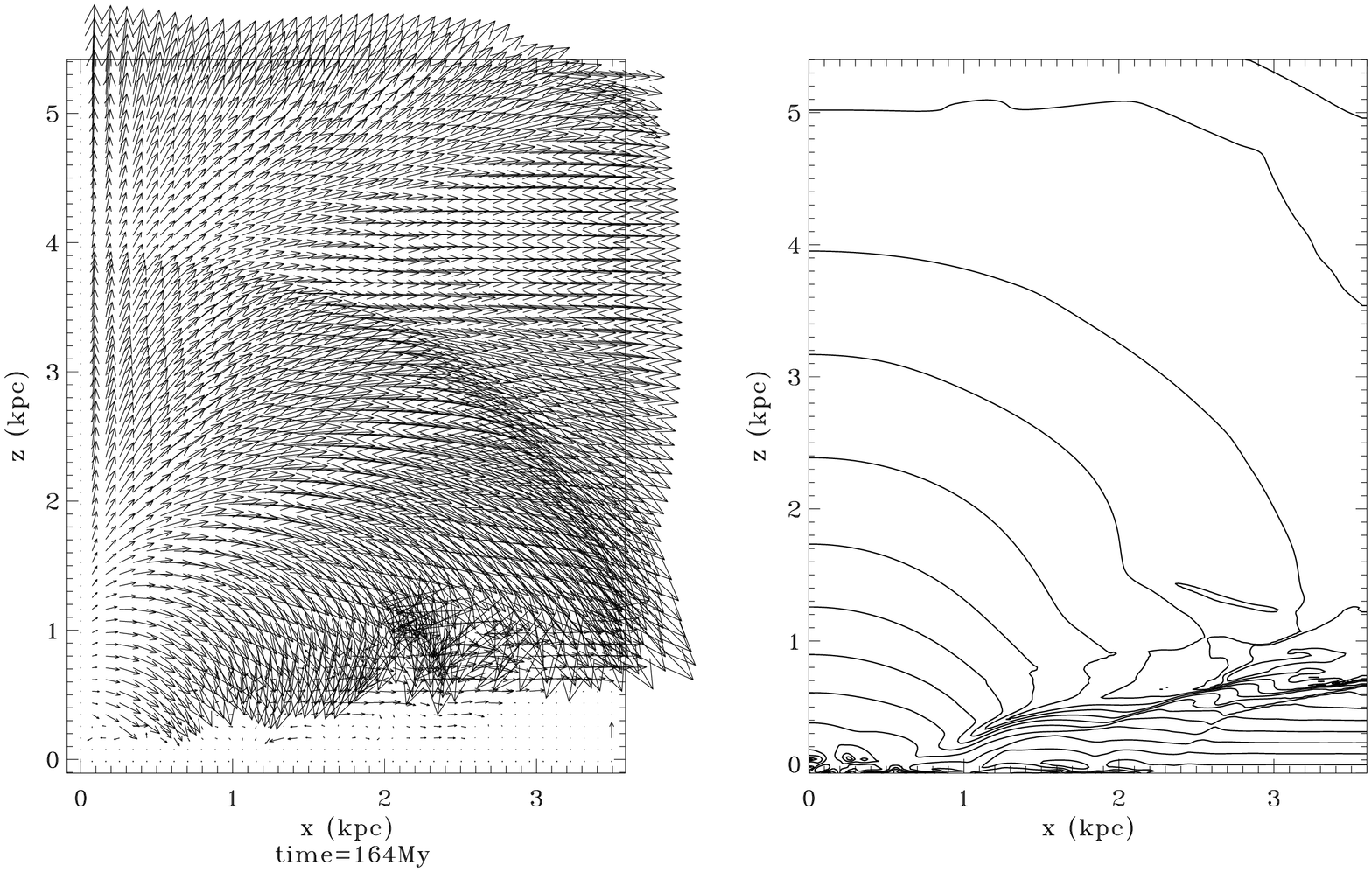,height=120mm,angle=90}}
 \caption{\footnotesize{Velocity vectors and magnetic field lines at
$t=41$ Myr (top) and $t=164$ Myr (bottom) for $g=6\times10^{-9}$ cm s$^{-2}$
and $\gamma=1.6$.}}
    \end{figure*}

For $\gamma=1.4$, the evolution is similar and the growth time is roughly the same. 
Subsequent time steps show the rising large-scale Parker loops reaching far out
into the halo up to $z=3$ kpc.
In addition, the
loops cause a radial outflow to be driven ahead.
For all $\gamma$, the largest vertical velocities
are determined by the gas sliding down along the footpoints of
the loops, and can reach $3c_s$ (where $c_s$ is the sound speed in the
disc, which we have represented in the lower right corner of the
left panel).  

In the linear theory, the Parker instability grows only when 
$\gamma$ is less than the critical value $\gamma_c=(1+\beta^{-1})^2/(1+
3\beta^{-1}/2)$, which for $\beta=1$ is 1.6. For larger $\gamma$, MHD 
waves disperse the perturbation on time scales which are less than the
free-fall time. When the disturbance is nonlinear, it can form and then 
maintain a localized structure which allows
gas to slide down along the curved magnetic lines, and thus even for 
$\gamma\geq\gamma_c$, one can expect the Parker instability to grow. 
In the above-described model 2Tg4.5k1.6, both $\lambda_c$ and 
$\lambda_m$ are indefinite. At the same time, the critical value for $\gamma$ has
been derived in the theory of linear perturbations. However, for nonlinear perturbations
the dispersion relation will change and $\gamma_c$ is expected to change as well.
Even in the  model 2Tg4.5k1.75
we observe formation of loops, but on larger time scales compared
to the previous two models. In Fig. 4, we show the velocity vectors and magnetic field
lines for $t=156$ Myr.

\subsubsection{Large $g$}

All three investigated models 2Tg6k1.05, 2Tg6k1.4 and 2Tg6k1.6 show vigourous growth
of the Parker instability on scales less than or comparable with the
rotation period of the galactic disc. 
In Fig. 5, we show the evolution of the instability for $\gamma=1.05$. 
The upper panel
shows the propagation of the perturbation at $t=41$ Myr, while the middle panel is already representing
the fully evolved instability after only 199 Myr. The main loop centred in the origin
has roughly the size of the wavelength of maximum growth (see table 2) while the secondary
loop already visible at this stage forms at about the critical wavelength.

For larger $\gamma$, the equivalent energy input is strong enough to 
lead to a supersonic shell when the front reaches the hot 
halo. This can be seen in the lower panel of Fig. 6. The front is pushing
a magnetosonic amplifying wave upwardly which soon becomes
nonlinear and forms a secondary shell. 
Compared to $\gamma=1.05$, the dynamics of the system is much
more vigorous. The developed magnetic field structures are
much more extended in $x$. 
We attribute this to the interaction
between the Parker flow and the blow-out and to the larger critical wavelength. 
Secondly, the magnetic field lines
show a more shell-like structure, indicating the presence
of a strong radial outflow, which can be seen in the upper panel
of Fig. 6 at $t=164$ Myr. 
The maximum velocity of the flows reaches about $6 c_s$.  

\section{Discussion}

In all cases we considered, the energy input $E$ is always less than the minimal 
energy $E_{\rm min}^B$ required for a blow out by a factor 2--5 depending on $g$ 
and $\gamma$, but larger than the minimal energy $E_{\rm min}^P$ needed 
for the Parker instability to grow on one rotational time $t_R$. 
We are thus in the range where the standard blow out does not operate, and 
the outflows produced by an explosion are connected with the growing Parker 
instability. 

The explosions with $E\simgt E_{\rm min}^P$ initiate 
the nonlinear Parker instability during the later stages of their evolution. 
The growth time related to the rotation time of the galactic disc depends
on the gravitational acceleration and the ratio of the specific heats. When the
characteristic dynamical time for a weakened shock wave and the growth time 
for the Parker instability are comparable, the two processes interact. 
Even though the shock wave is transsonic (Mach number 
$M\sim 1$), it produces a nonlinear perturbation in the velocity and 
magnetic field, $v\sim c_s$ and $\delta B\sim B_0$, respectively. 
At these conditions the compression of gas and magnetic field and 
the Parker flow 
operate simultaneously. The Parker instability 
evacuates 
gas from upper parts of the magnetic loop, resulting in a faster expansion 
of the interior hot bubble, or the expanding bubble is carrying
material away, allowing for the magnetic field to rise more easily and for
the instability to grow faster. Due to this interaction, the shell looses
mass.
As a consequence, the total amount of gas which can 
be ejected into the halo is expected to be less than in cases when the Parker instability 
does not operate. At the same time, the magnetic loops formed by such flows
must be more prominent and must extend further out into the halo. From this
point of view, the structures built differ qualitatively from those which arise
when the Parker instability is initited by a 
non-localized linear sinusoidal 
perturbation. In the latter case, a steady state appears to be reached and the loops
extend up to 
$\sim 1$ kpc with respect to the plane (Kim \etal 2000). 
We can thus conclude that the origin of the large scale 
magnetic loops observed in haloes of edge-on galaxies 
might 
likely be connected with explosions in the underlying discs with the explosion energy 
lying in the range $E_{\rm min}^P<E<E_{\rm min}^B$. 
Assuming the luminosity 
function of OB associations to be universal (a power-law, Williams \& McKee, 
1997) one can expect that large 
scale magnetic arches must be more numerous in galaxies with larger 
gravitational acceleration $g$, as both $E_{\rm min}^P$ and $E{\rm min}^B$ 
decrease with $g$. 

Although we have not considered the limit $E>E_{\rm min}^B$, we
expect the hot bubble and the shell to 
expand much faster than the Parker instability evolves, so that 
the effects of removal of gas from the shell can be neglected, and the total 
amount of mass ejected into the halo is comparable to that expected 
for explosions in a non-magnetic ISM.  Such explosions also produce large 
scale magnetic arches in galactic halos. However, in this case, a much larger
amount of gas must be associated with the loops in comparison with 
explosions having smaller energies. It is evident that magnetic 
structures of this kind must be less numerous in galactic halos since 
$E_{\rm min}^P<E_{\rm min}^B$.

We conclude that the ejection of mass and magnetic fields into the galactic 
haloes due to blowing out SNe explosions is most efficient in galaxies with 
large gravitational acceleration $g$. Consequently, galaxies with large 
$g$ must have more massive halos (relative to the mass of their 
gaseous discs), smaller scale heights, and more prominent 
large scale magnetic arches. Assuming that the gas mass in galactic halos 
is proportional to the number of OB associations with the SNe energy 
larger than $E_{\rm min}^B$, and that the luminosity function of associations 
is universal, one can expect the mass of the halo gas to vary as 
proportional to $g^3$. 
By comparing the energy limits 
$E_{\rm min}^P$ and $E_{\rm min}^B$, one can conclude that the large scale magnetic Parker 
loops in the halo of a given galaxy must be more 
numerous than the blow-out events. From the observational point of view, these  
structures differ from each other mainly by their plasma $\beta$ parameter: 
in blow-out flows $\beta\sim 1$ since the frozen-in magnetic field is  
compressed by a gas flow that is predominantly transverse, while in magnetic loops 
formed due to the Parker instability,
$\beta\ll 1$, since the gas flow here is predominantly longitudinal (Mouschovias 
1974). However, in order to distinguish these
structures, one needs to measure both the strength of the magnetic field
and the gas pressure in galactic halos, which is beyond the present
observational possibilities.
 
In all cases, we assumed equipartition between thermal and magnetic pressure
for the unperturbed background state.
However, both the critical wavelength
and the characteristic growth rate for the Parker instability depend on $\beta$:
for a fixed $\gamma$ the critical wavelength becomes
infinite and the instability does not develop when $\beta$ is larger
than a critical value $\beta_c(\gamma)$. The growth rate decreases
in the limits $\beta<1$ and $\beta>1$ and reaches a maximum around $\beta\sim 1$.
One can therefore expect that in galaxies where the equipartition between
thermal and magnetic pressures is not given,
the Parker instability is supressed, and only blow-out flows
can occur, provided the energy input from SNe explosions is large enough
$E>E_{\rm min}^B$.

All the above conclusions have been drawn from 2D studies of the
Parker instability. In a 3D treatment, the growth of small wavelength
perturbations due to the manifestation of the interchange mode may
change the dynamics of the system. Kim et al. (1998) have shown that
the nonlinear phase and the final structures of the Parker instability
in 3D can be different from those in 2D simulations. However, one has
to bear in mind that they have assumed
equipartition between thermal and magnetic pressure in the initial state and
the linear growth of the Parker instability in 3-D, when the magnetic pressure
is large compared to the thermal pressure is very different from the situation
when beta is uniy. 
Based on this later assumption, they have demonstrated
that in the planes perpendicular to the direction of the magnetic and
gravitational field, the flow can be characterized as chaotic, while
in the plane that contains the magnetic and gravitational field, the
structures produced by the undular mode are still present  
and persist over a period of about 40 $H/c$. This time translates
into about 400 Myr in our model for $g=6 \times 10^{-9}cm/s^2$ an would
leave enough time for the interaction we are concerned with to
operate in 3D as well. Apart from this aspect, we study the evolution
of the instability under a strongly nonlinear perturbation, the SN explosion.
Compared to randomly applied perturbations, which are associated with
equally distributed power in all possible wavenumbers, this perturbation has
most of its power in the low wavenumber modes. The large wavenumbers favoured
by the interchange mode will grow as well, but it will take longer for
them to reach dynamically significant amplitudes. In addition they will
be masked by the supersonic motions associated with the SN explosion
and later the supersonic motions of the undular mode. Since our system
is more complicated than the usual picture in which linear perturbations
are used, it is not possible to completely infer the evolution of
such a system from purely intuitive arguments. 3D simulations of the
SN explosion induced Parker instability will provide us with the final
answer and we will address this issue in a forthcoming work.

\section{Summary}
In this paper our primary interest has focused on the evolution of
the Parker instability induced by a single SN explosion.
We have explored the dependence of
the instability on the gravitational acceleration in the host galaxy and
the ratio of specific heats, while we kept the energy input of the explosions
as well as the volume in which this energy was distributed constant.  
It was our aim to address the question of how the energy input from 
an explosion converts into large scale motions corresponding to the Parker flow,
depending on the gravitational acceleration in a host galaxy. Our results can be 
summarised as follows: 

1. The question of whether the Parker instability triggered by a single
SN explosion can lead to the observed
large scale magnetic field loops in the halos of galaxies
is strongly connected to the strength of the gravitational
acceleration in the host galaxy, the energy of the explosion and
the ratio of specific heats. We found that for values of the gravitational
acceleration less than $4.5\times 10^{-9}$ cm s$^{-2}$, the instability
is growing on much larger time scales in comparison to the models where
a sinusoidal perturbation was applied in order to trigger the instability. 
The relevance of the instability in this case must be judged by relating
its growth time to other dynamically important time scales (e.g. the
rotation time of the galactic disc at the given distance from its centre).  
For larger $g$ (around $6\times 10^{-9}$ cm s$^{-2}$) the growth
rates for the instability are comparable to the models with nonlocalised
perturbations. In this case, we obtained large growth rates for the instability.
We conclude that in galaxies with large $g$ in the disc, the Parker instability
is relevant on dynamical time scales. 

2. The choice of $\gamma>1$ is equivalent to a larger amplitude of the
localised perturbation and leads to larger growth rates of the instability.
We obtained significant growth even for $\gamma$ larger than the critical
value $\gamma_c=1.6$ known from the linear theory. This result is
in agreement with the results obtained by Kamaya et al. (1996). One reason
for this apparently contradictory behaviour is the applied nonlinear  
perturbation which changes the dispersion relation, and therefore the growth
rate of the instability. The linear critical $\gamma$ therefore should
no longer be valid in this case. On the other hand,
strong nonlinear perturbations are unaffected by the dispersive properties of
the background medium.

3. Even one SN explosion can lead to the formation of multiple
loops due to the fact that the perturbation propagates
in the disc. The secondary loops in general are associated with
smaller growth rates.

4. The SN explosion and the Parker instability are two dynamical
processes that can interact and can influence each other depending
on the given energetics, \ie the minimal energy $E^P$ required
for the Parker instability to be initiated and the minimal
energy $E^B$ for the blow-out. In general, we distinguish three
cases: a) If $E^P < E << E^B$ the observed magnetic structures
are a consequence of the Parker instability; b) For $E^P < E\simlt E^B$ 
the two processes interact and the resulting structures are a consequence
of this interaction; c) For $E>E^B$ the ejection of mass and magnetic field
into the halo is mainly due to the blow-out flow. Our simulations are only
adressing case b). Observed outflow rates should reflect these
three regimes. The dependence of the dynamics of the system on $g$
can be converted into the dependence on the input energy via
the scaling laws $Eg^2=$const in the 2D case and $Eg^3$=const in 3D. 

\vskip 1truecm

We thank D. Bomans, R.-J. Dettmar, V. Korchagin and A. Schr\"oer 
for valuable discussions and the referee, Tom Hardquist, for his
careful reading of the manuscript and critical comments.
The authors acknowledge 
financial support from the German Science Foundation (DFG)
within Sonderforschungsbereich 191.



\vskip 2truecm
\begin{thebibliography}{99}

\bibitem{} Appenzeller, I. 1974, A \& A, 36, 99 

\bibitem{} Bahcall, J. N. 1984, ApJ, 276, 169

\bibitem{b97}Basu, S., Mouschovias, T. Ch., Paleologou, E. V. 1997, ApJ, 
480, L55

\bibitem{} Bernstein, I. B., Kulsrud, R. M. 1965, ApJ, 142, 479

\bibitem{} Bottema, R. 1995, The stellar kinematics of galactic disks, 
PhD thesis, Rijksuniversiteit Groningen 

\bibitem{} Brinks, E., Bajaja, E. 1986, A \& A, 169, 14

\bibitem{} Dettmar, R.-J. 1992, Fund. Cosmic Phys., 15, 143 

\bibitem{} Dickey J. M., Lockman F. J., 1990, ARA\& A, 28, 215

\bibitem{} Duric, N., Seaquist, E. R., Crane, P. C., Bignell, R. C., 
Davis, L. E. 1983, ApJ, 273, L11

\bibitem{} Ferrara, A., Tolstoy, E. 2000, MNRAS, 313, 291

\bibitem{} Ferri\'ere, K. M., Mac Low, M.-M., Zweibel, E. 1991, ApJ, 375, 239  

\bibitem{} Giuliani, J. N. 1982, ApJ, 256, 624

\bibitem{g93}Giz, A. T., Shu, F. H. 1993, ApJ, 404, 185 

\bibitem{h84}Heiles, C. 1984, ApJS, 55, 585

\bibitem{h88}Horiuchi, T., Matsumoto, R., Hanawa, T., Shibata, K. 1988, PASJ, 40, 147

\bibitem{h97}Howk, J. C., Savage, B. D. 1997, AJ, 114, 2463

\bibitem{} Howk, J.C., Savage, B.D. 1999, AJ 117, 2077

\bibitem{} Hummel, E., van Gorkom, J. H., Kotanyi, G. G. 1983, ApJ, 267, L5

\bibitem{} Kalberla P. M. W., Kerp J., 1998, A\&A, 339, 745

\bibitem{k97}Kamaya, H., Horiuchi, T., Matsumoto, R., Hanawa, T., 
Shibata, K., Mineshige, S. 1997, ApJ, 486, 307

\bibitem{k96}Kamaya, H., Mineshige, S., Shibata, K., Matsumoto, R. 1996, ApJ, 458, L25

\bibitem{k99}Kim, J., Franco, J., Hong, S. S., Santillan, A., Martos, M. A. 
2000, ApJ, 531, 873

\bibitem{} Kim, J., Hong, S. S. 1998, ApJ, 507, 254

\bibitem{ki97}Kim, J., Hong, S. S., Ryu, D. 1997, ApJ, 485, 228

\bibitem{} Kim, J., Hong, S. S., Ryu, D., Jones, T. W. 1998, ApJ, 506, L139

\bibitem{} Kim, J., Ryu, D., Jones, T. W., Hong, S. S. 1999, ApJ, 514,
506

\bibitem{} Kim, S., Dopita, M. A., Staveley-Smith, L., 
Bessel, M. S. 1999, AJ, 118, 2797

\bibitem{} Kovalenko, I. G., Shchekinov, Yu. A. 1985, SovA, Astrophysics, 1985, 
23, 578

\bibitem{} Kuijken, K., Gilmore, G. 1989, MNRAS, 239, 605

\bibitem{} Kulsrud, R. M., Bernstein, I. B., Kruskal, M., Fanucci, J., 
Ness, N.  1965, ApJ, 142, 491

\bibitem{} Lamb, H. 1932, Hydrodynamics, CUP, Cambridge

\bibitem{} Mac Low, M.-M., Ferrara, A. 1999, ApJ, 513, 142 

\bibitem{} Mac Low, M.-M., McCray, R. 1988, ApJ, 324, 776

\bibitem{} Mac Low, M.-M., McCray, R., Norman, M. 1989, ApJ, 337, 776

\bibitem{} Martin, C. L. 1996, ApJ, 465, 680

\bibitem{} Matsumoto, R., Horiuchi, T., Shibata, K, Hanawa, T. 1988, 
PASJ, 40, 171 

\bibitem{} Matsumoto, R., Tajima, T., Shibata, K., Kaising, M. 1993, 
ApJ, 414, 357 

\bibitem{} Meurer, G. R., Freeman, K. C., Dopita, M. A., Cacciari, C. 1992, A
J, 103, 60

\bibitem{} Mineshige, S., Shibata, K., Shapiro, P. R. 1993, ApJ, 409, 663

\bibitem{} Mouschovias, T. 1974, ApJ, 192, 37 

\bibitem{m96}Mouschovias, T. 1996, in Solar and Astrophysical Magnetohydrodynamic Flows,
Kluwer Academic Publishers, Dordrecht, p. 475 

\bibitem{} Newcomb, W. A. 1961, Phys. Fluids, 4, 391 

\bibitem{} Oort, J. H. 1960, Bull. Astr. Inst. Netherlands, 15, 45 B

\bibitem{} Palou\v s, J., Franco, J., Tenorio-Tagle, G. 1990, A \& A, 227, 175

\bibitem{p66}Parker, E. N. 1966, ApJ, 145, 811

\bibitem{p66}Parker, E. N. 1967, ApJ, 149, 535

\bibitem{p79}Parker, E. N. 1979, Cosmic Magnetic Fields, Oxford University Press, 
Oxford 

\bibitem{} Puche, D., Westpfahl, D., Brinks, E., Roy, J.-R. 1992, 
AJ, 103, 1841 

\bibitem{} Scalo, J. M. 1985, in: Protostars and Planets II, Tucson, AZ, 
University of Arizona Press, p. 201  

\bibitem{} Shibata, K., 1996, in Solar and Astrophysical
Magnetohydrodynamic Flows, Kluwer Academic Publishers, Dordrecht, p. 232

\bibitem{s89}Shibata, K., Tajima, T., Matsumoto, R., Horiuchi, T., Hanawa, T., 
Rosner, R., Uchida, Y. 1989, ApJ, 338, 471

\bibitem{s76}Sofue, Y., 1976, A\& A, 48, 1

\bibitem{} Sofue, Y., Handa, T. 1984, Nature, 310, 568 

\bibitem{} Sofue, Y., Tosa, M. 1974, A \& A, 36, 237

\bibitem{s96}Sofue, Y., Wakamatsu, K.-I., Malin, D. F. 1994, AJ 108, 2102

\bibitem{} Stone, J.M., Norman, M. L. 1992a, ApJS, 80, 791

\bibitem{} Stone, J.M., Norman, M. L. 1992b, ApJS, 80, 819

\bibitem{} Tenorio-Tagle, G., Bodenheimer, P., Rozyczka, M. 1987, A \& A, 
182, 120 

\bibitem{} Tomisaka, K. 1990, ApJ, 361, L5

\bibitem{} Tomisaka, K. 1998, MNRAS, 298, 797

\bibitem{} Tomosaka, K., Ikeuchi, S. 1986, PASJ, 38, 697 

\bibitem{}T\"ullmann, R., Dettmar, R.-J. 2000, A\& A, 362, 119

\bibitem{} van der Kruit, P. C., Searl, L. 1981, A\& A, 95, 105

\bibitem{} Walter, F., Brinks, E. 1999, AJ, 118, 273

\bibitem{} Williams J. P., McKee, C. F. 1997, ApJ, 476, 144

\bibitem{} Ziegler, U., 1995, PhD, University of Wuerzburg

\bibitem{} Ziegler, U., R\"udiger, G. 2000, A \& A, 356, 1141

\end{thebibliography}
\end{document}